# Inferring agent objectives at different scales of a complex adaptive system

**D. Hendricks**[*]  **A. Cobb**  **R. Everett**  **J. Downing**  **S.J. Roberts**

Machine Learning Research Group
Oxford-Man Institute of Quantitative Finance
University of Oxford
[*]Corresponding author: `dieter.hendricks@eng.ox.ac.uk`

## Abstract

We introduce a framework to study the effective objectives at different time scales of financial market microstructure. The financial market can be regarded as a complex adaptive system, where purposeful agents collectively and simultaneously create and perceive their environment as they interact with it. It has been suggested that multiple agent classes operate in this system, with a non-trivial hierarchy of top-down and bottom-up causation classes with different effective models governing each level. We conjecture that agent classes may in fact operate at different *time scales* and thus act differently in response to the same perceived market state. Given scale-specific temporal state trajectories and action sequences estimated from aggregate market behaviour, we use Inverse Reinforcement Learning to compute the effective reward function for the aggregate agent class at each scale, allowing us to assess the relative *attractiveness* of feature vectors across different scales. Differences in reward functions for feature vectors may indicate different objectives of market participants, which could assist in finding the scale boundary for agent classes. This has implications for learning algorithms operating in this domain.

## 1  Introduction

Equity financial markets consist of multiple competing agents operating through a centralised electronic exchange, giving rise to non-linear interactions both through time and among agent classes. *Investors* use their understanding of asset dynamics to time buying and selling decisions for financial gain, *traders* use their understanding of market dynamics to plan trades and minimise the cost of realising investment decisions and *market makers* use their understanding of investor demand to profit from liquidity provision. The field of market microstructure [1] studies the dynamics of price formation in this system at intraday time scales, considering how mechanistic rules, regulatory oversight and social behaviours of participants interact to manifest the observed time series'. The *Complexity Economics* [2] paradigm seeks to explain observed behaviours through the lens of complex adaptive systems, where competing agents continually adapt their actions and strategies based on the observed state they mutually create. Wilcox and Gebbie [3] take this further, proposing a mechanism for bottom-up and top-down causation, with level-specific effective models governing actors and inter-level interaction via noise terms. Actors at each level perceive the system in a different way, which invalidates the use of hierarchies of the same effective model to capture system complexity.

We are interested in developing learning algorithms in this domain, where an understanding of scale-specific state representation, in the context of competing agents, is key to ensuring relevant features can be exploited and useful learning can take place faster than the natural time-scale of the system. Galla, Farmer and Sanders [4, 5] investigate the nature of agent learning in complicated games, using Experience Weighted Attraction (EWA) to evaluate the propensity for asymptotic learning. They demonstrate the importance of understanding agent payoff (reward) correlations when setting learning rates, to ensure that learning is feasible and chaotic regimes are avoided. This study seeks to understand the landscape of competing agents at different time scales, to interrogate the importance of scale for learning policies in this domain.

We use Inverse Reinforcement Learning (IRL) to compute the effective reward function at different scales of equity market microstructure, using scale-specific temporal state trajectories and action sequences estimated from aggregate market behaviour. This allows us to identify *attractive* states and assess the attractiveness of the associated feature vector across different scales. This is a first step towards understanding relative objectives of agent classes at different scales of this system.

## 2 Method

### 2.1 Inverse Reinforcement Learning for state attractiveness

IRL aims to infer the reward function in a Markov decision process (MDP) defined by the tuple $m = \langle \mathcal{S}, \mathcal{A}, \mathcal{P}, \gamma, \mathcal{R} \rangle$, describing the state space $\mathcal{S}$, the action space $\mathcal{A}$, the transition function $\mathcal{P}$, the discount factor $\gamma$ and the reward $\mathcal{R}$. Given access to sample state-action space trajectories collected from observing an agent's behaviour, the objective is to find a reward function that induces agents to follow trajectories matching the expert trajectories.

While many IRL algorithms have been proposed ([6, 7, 8, 9]), we will use Maximum entropy (MaxEnt) IRL [10], which represents a move towards looking at probability distributions over paths. Ziebart et al. employ the principle of maximum entropy to pick the least informative set of parameters for a linear reward function $r = \theta^\top \mathbf{f}_\zeta$, a linear combination of state-action trajectories $\mathbf{f}_\zeta$ that match the feature expectations between an agent's observed trajectories and the learner's behaviour.

This matching equation,

$$\sum_{\text{Path } \zeta_i} P(\zeta_i) \mathbf{f}_{\zeta_i} = \tilde{\mathbf{f}} \tag{1}$$

provides the constraints for maximum entropy, where $\tilde{\mathbf{f}} = \frac{1}{m} \sum_i \mathbf{f}_{\tilde{\zeta}_i}$ is the average empirical feature count from $m$ observed trajectories. Therefore summing the feature mappings along a trajectory results in $\mathbf{f}_{\zeta_i} = \sum_{s_j \in \zeta} \mathbf{f}_{s_j}$.

Ziebart et al. evaluate a probability distribution over trajectories through space $P(\zeta_i \mid \boldsymbol{\theta}, T)$ that give trajectories with equal rewards, equal probabilities and assign an exponentially higher preference to higher rewards. Note that $T$ is the transition model of the MDP. The optimal set of parameters $\boldsymbol{\theta}^*$ is then obtained by maximising the likelihood, $L(\boldsymbol{\theta})$, i.e.

$$\boldsymbol{\theta}^* = \underset{\boldsymbol{\theta}}{\operatorname{argmax}}\, L(\boldsymbol{\theta}) = \underset{\boldsymbol{\theta}}{\operatorname{argmax}} \sum_{\text{examples}} \log P(\tilde{\zeta} \mid \boldsymbol{\theta}, T) \tag{2}$$

Since this function is convex for deterministic MDPs, we can use gradient optimisation methods, requiring knowledge of the gradient given as

$$\nabla L(\boldsymbol{\theta}) = \tilde{\mathbf{f}} - \sum_\zeta P(\tilde{\zeta} \mid \boldsymbol{\theta}, T) \mathbf{f}_\zeta = \tilde{\mathbf{f}} - \sum_{s_i} D_{s_i} \mathbf{f}_{s_i}. \tag{3}$$

This gradient represents the difference between the empirical expected feature counts from the observations and the learner's expected feature counts. The difficulty in the optimisation process is in computing the value of the *expected state visitation frequencies*, $D_{s_i}$ to calculate the gradient. The details of the algorithm for $D_{s_i}$ are given in the Ziebart et al. [10]. To summarise, it involves a backwards pass to calculate $P(\zeta_i \mid \boldsymbol{\theta}, T)$ and then followed by a forward pass to calculate the expected state visitation frequencies. Note that a large horizon is used to calculate the state frequencies to approximate the infinite time horizon of the MDP.

We will use the MaxEnt IRL implementation provided by Matthew Alger [11].

### 2.2 Determining temporal state trajectories

Hendricks et al. [12] propose an approach for the detection and online estimation of intraday temporal states from equity market microstructure features. They found an interesting hierarchy of system behaviour at different calendar time scales, with results suggesting there may be different universality classes characterising behaviour at each scale. The approach uses an analogy to the q-state Potts model to develop an unsupervised clustering technique consistent with finding meta-stable object configurations in complex systems. Time periods are clustered into states based on observed *trade price, spread, volume* and *volume imbalance* features for major stocks on an exchange. Significant states are identified, with the associated State Signature Vectors (SSV) used for online state detection and assignment. Figure 1 illustrates the temporal states and associated feature vectors for 60-minute and 30-minute time scales. Each node represents a time period within the month, with node shading indicating the time of day and node connectedness indicating cluster (state) membership.

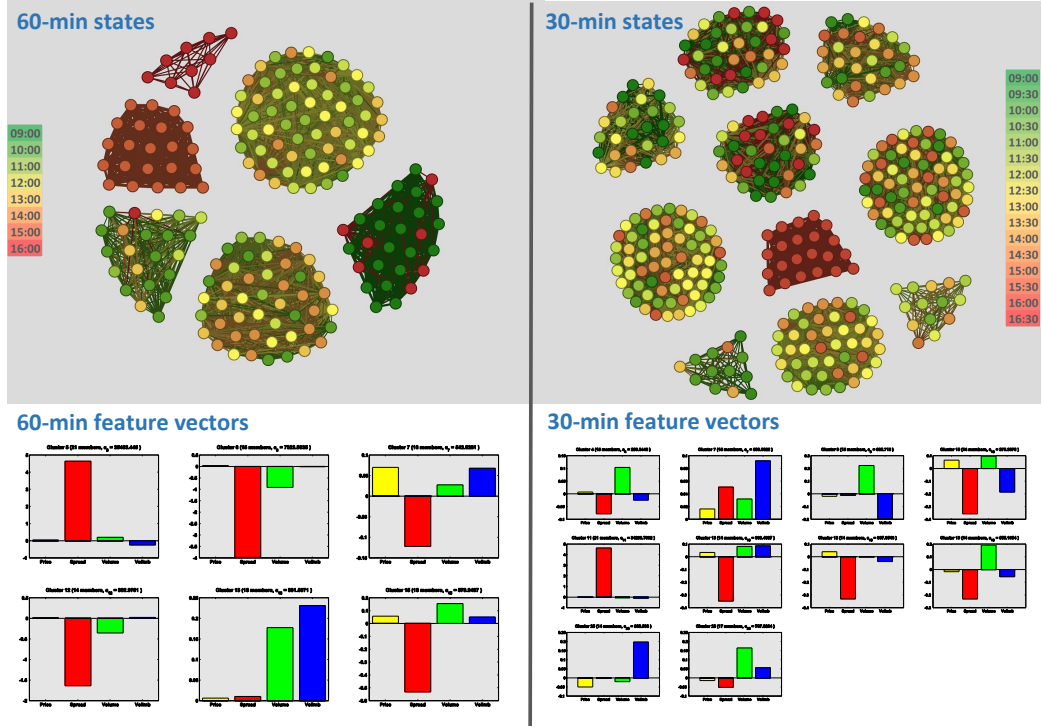

Figure 1: (Above) Each node represents a time period, with the colour shading indicating the time-of-day (Morning = green, Lunch = yellow, Afternoon = red) and node connectedness indicating identified states. (Below) Average change in trade price, spread, trade volume and quote volume imbalance across member periods in each temporal state.

We will construct temporal state trajectories at different calendar scales based on the SSV state assignment using the method in [12]. The chronological sequence of cluster membership indices thus defines the associated state trajectory at each scale.

Given the objective of categorising agent *classes* at different scales, we will use the sequence of *average price returns* at each scale to determine associated actions. The motivation is that if the behaviour at a certain scale resulted in a *positive (negative) price return*, this could indicate *net buying (selling)* decisions for that class. We will thus use the sign of average price return to assign the action to one of three states, {*BUY, SELL, NEUTRAL*}.

## 3 Experiments

**Data** State and action trajectories were computed from tick-level trades and top-of-book quotes for 42 stocks on the Johannesburg Stock Exchange (JSE) from 1 November 2012 to 30 November 2012, sourced from the Thomson Reuters Tick History (TRTH) database. This data was resampled based on the scale considered. We use four features for each stock: Change in i) trade *price*, ii) quote *spread* (Ask price - Bid price), iii) trade *volume* and iv) quote *volume imbalance* ($\frac{\text{Ask volume}}{\text{Bid volume}}$).

**Results** Figure 2 shows the feature vectors and estimated reward functions for each state at the four candidate time scales: 5, 15, 30 and 60 minutes. It is interesting that low *spread* states are unfavourable, perhaps because there are fewer opportunities when markets are highly liquid, whereas high *spread* states provide opportunity for profit via market dislocations. Upon examination of the relative rewards and feature vectors, it appears there are differences in attractiveness across scales, although plausible explanations for these results are difficult without knowledge of the ground truth.

Figure 3 shows the relative reward across scales for similar feature vectors. We used a simple $K$-means clustering algorithm to group *all* states into 6 clusters on the basis of feature vector similarity. We then sum the (normalised) rewards for all states at the same scale in each cluster. This allows us to consider the net reward for each scale given the same feature vector. Node size indicates scale (60-min=large, 5-min=small) and colour indicates reward (red=negative, green=positive).

Clusters 1-5 all show both positive and negative rewards, providing some evidence that agents react differently to observed features at different scales. Interpreting the rewards using relative cluster centroid *spread* and *volume imbalance* provides some intuition.

For Cluster 1, negative *volume imbalance* and negative *spread* could indicate a larger quantity of aggressive buy orders, which could have a short-term negative effect of buying at a higher price, but if the stock is held at longer time scales, there could be a net gain in asset value.

For Cluster 3, we see a larger negative *volume imbalance* and negative *spread*, which appears to be favourable at the 5-min scale, but unfavourable at longer time-scales. This suggests that the *magnitude* of certain features may affect the state's attractiveness, at least when measured using aggregate buying/selling behaviour as actions.

For Cluster 5, positive *volume imbalance* and negative *spread* indicate a larger quantity of aggressive sell orders. This could be positive at the 5-min scale, where a short-term sell is more easily matched, but the increase in aggressive ask quotes could translate into downward price pressure, leading to negative rewards at higher scales.

For Cluster 6, negative *volume imbalance* and positive *spread* could be enable agents to buy larger quantities of stocks at lower prices, with the imbalance leading to further price appreciation across all scales.

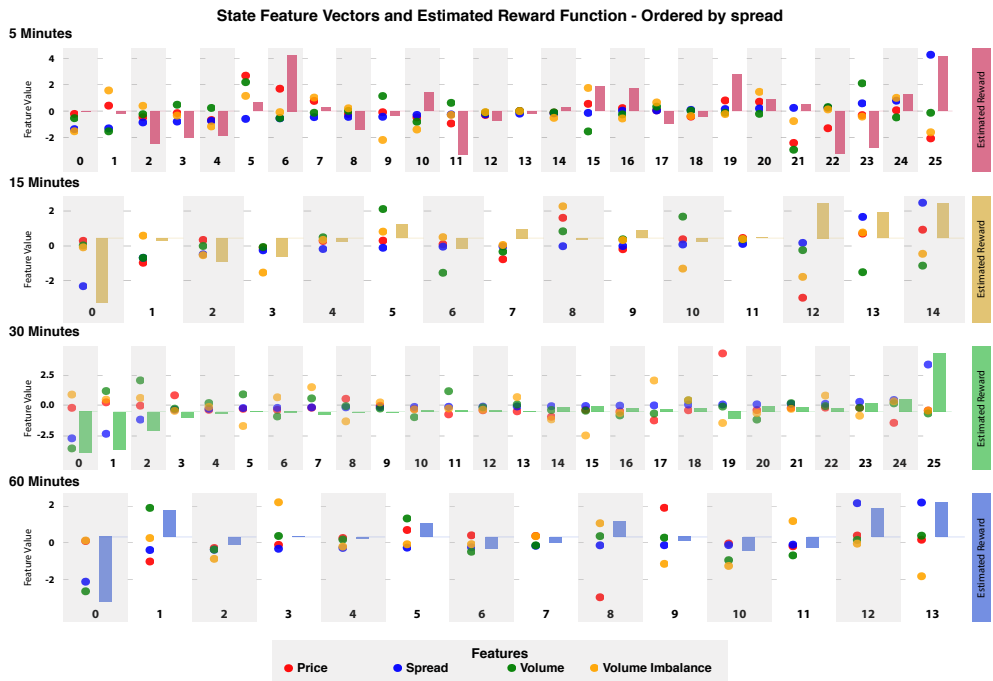

Figure 2: This figure shows the feature vectors (dots) and estimated reward function (bars) for each state, at 4 candidate time scales (5, 15, 30 and 60 minutes)

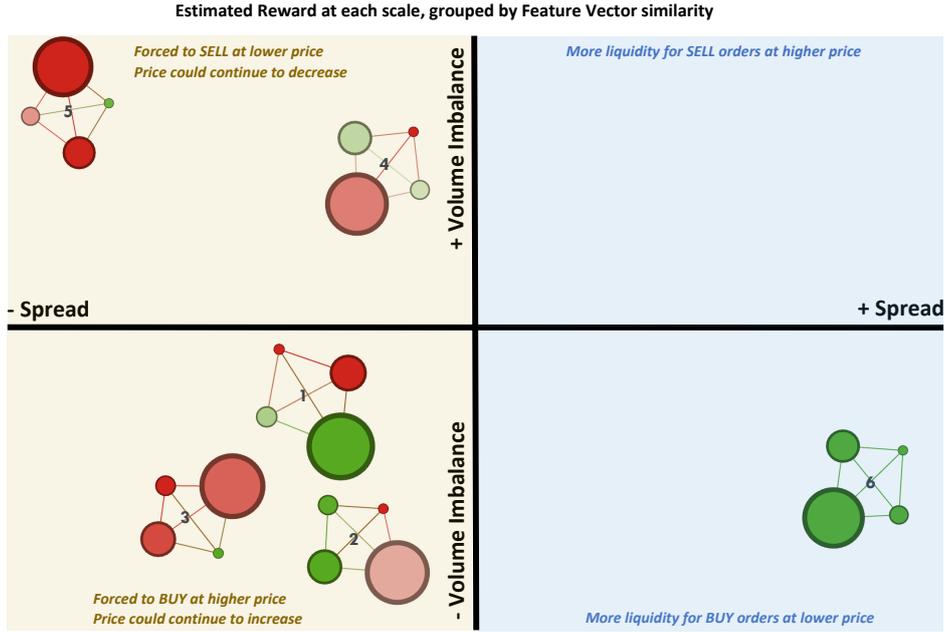

Figure 3: This figure illustrates the relative reward across scales for similar feature vectors. Node size indicates scale (smallest=5min, largest=60min), node colour indicates reward (red=low negative, green=high positive), nodes are clustered by feature vector similarity, and clusters are positioned by the relative *spread* and *volume imbalance* feature values of the cluster centroids.

## 4  Discussion and future work

We have provided a framework which allows us to interrogate the relative attractiveness of feature vectors at different scales of a complex adaptive system. While we have suggested plausible explanations for observed feature vector attractiveness, more work is required to analyse the inducing policies from estimated reward functions and assess whether this suggests particular objectives or adversarial behaviour across scales. Ultimately, this approach can help to assess whether a scale boundary exists for agent classes, to better inform learning algorithm specification. This could lead to a hierarchical reinforcement learning framework with multi-scale learning in financial markets to exploit hierarchy of causality at different scales. Further work will assess a complete spectrum of event time scales, consistent with how machine algorithms process information, to provide clarity on boundaries between high-frequency (algorithmic) market making, machine trading, human trading and investment decisions.